\documentclass[a4paper]{jpconf}
\usepackage{graphicx}
\usepackage[numbers,square,sort&compress]{natbib}
\begin{document}
\title{Dynamos in Stellar Convection Zones: of Wreaths and
  Cycles\footnote{proceedings for SOHO 24/GONG 2010 conference: ``A
    new era of seismology of the Sun and solar-like stars,''
    Aix-en-Provence, France, June 27-July 4, 2010 (JPCS)}}

\author{Benjamin Brown}

\address{Dept. of Astronomy and Center for Magnetic Self Organization
  in Laboratory and Astrophysical Plasmas, University of Wisconsin,
  Madison, WI 53706, USA}

\ead{bpbrown@astro.wisc.edu}

\begin{abstract}
We live near a magnetic star whose cycles of activity are driven by
dynamo action beneath the surface.  In the solar convection zone,
rotation couples with plasma motions to build highly organized
magnetic fields that erupt at the surface and undergo relatively
regular cycles of polarity reversal.  Despite our proximity to the
Sun, the nature of its dynamo remains elusive, but observations of
other solar-type stars show that surface magnetism is a nearly
ubiquitous feature. In recent time, numerical simulations of
convection and dynamo action have taken tremendous strides forward.
Global-scale organization and cyclic magnetism are being achieved by
several groups in distinctly different solar and stellar simulations.
Here I will talk about advances on the numerical front including
wreath-building dynamos that may occupy stellar convection zones.
I will discuss the interplay between the new simulations, various
classes of mean-field models, and current and upcoming solar and
stellar observations.
\end{abstract}

\vspace{-1cm}
\section{Introduction}

The 22-year solar activity cycle stands out as one of the most
remarkable and enigmatic examples of magnetic self-organization in
nature.  The magnetism we see at the surface as sunspots likely
originates in the solar convection zone, where turbulent plasma motions
couple with rotation and magnetic fields to drive strong dynamo
action.  Magnetism is a ubiquitous feature of solar-type stars and
there are well known scaling relationships between the amount of
surface magnetism and stellar properties, such as rotation rate and
stellar-type \citep[e.g.,][]{Pizzolato_et_al_2003}.  Other late-type
stars undergo similar magnetic cycles, with periods ranging from
several years to several decades
\citep{Baliunas_et_al_1995,Olah_et_al_2009}.  As in the case of the
solar cycle, these must arise through hydromagnetic dynamos operating
in their convective envelopes.  However, the observational landscape
is complex, with few well-established trends to constrain dynamo
models \citep[e.g.,][]{Rempel_2008,Lanza_2010}.

Indeed, explaining the origin of globally organized fields and cyclic
behavior in the Sun has inspired and challenged astrophysical dynamo
theory for over a century and continues to do so.  The tremendous
growth of computational resources, coupled with the insights afforded
by helioseismology about the solar internal structure and
differential rotation, has lead to an explosion of dynamo modeling
efforts.  These range from sophisticated 2D mean-field models that
incorporate data assimilation techniques to fully 3D simulations that
can capture non-linear dynamics of solar convection and
self-consistently establish solar-like differential rotation profiles.
Both classes of models are being applied to stars other than the Sun,
sampling different spectral types and exploring how rotation affects
stellar convection and dynamo action.

Here I will briefly review the state of dynamos for the Sun and
sun-like stars that rotate somewhat faster, as the Sun did in its
youth.  These studies are suggesting new modes of global-scale dynamo
action and are raising exciting questions about the nature of the
solar dynamo.

\section{Mean-field models of stellar dynamos}

Simulations of the solar dynamo and solar convection are generally
broken into two classes: the two dimensional (2D) mean-field models and
the three-dimensional (3D) simulations.  By going to 2D, typically in radius and latitude, the computations can be made much more
tractable and high resolution simulations can be pursued for very long
intervals of time.  It is not unusual for such models to simulate
several tens of solar cycles, and reproducing the entire observational
record of sunspots is quite feasible
\citep[e.g.,][]{Dikpati&Gilman_2006}.

The trouble lies in the treatment of turbulence and non-linear
correlations.  Dynamo action generally requires correlations between
the non-axisymmetric, fluctuating velocity and magnetic fields.  In
mean-field models these non-axisymmetric flows are not directly
simulated and instead must be captured through some assumed model.
This is a very difficult and outstanding problem in turbulence theory.
Generally, an attempt is made to model the coherent effect of the
fluctuations in terms of the global-scale (mean) fields, and this is
often embodied as an ``$\alpha$-effect,'' though many variations have
been explored.  Cyclonic convection coupling with rotation is often
associated with the $\alpha$-effect and thus $\alpha$ is thought
typically to depend on the kinetic helicity of the convection, but the
$\alpha$-effect remains very difficult to constrain observationally.
Comparable difficulties underlie descriptions of the turbulent
processes that transport angular momentum to self-consistently
establish the observed solar differential rotation.  This mean
internal rotation profile, however, can be measured in detail
throughout the solar convection zone, and the observed profile is
directly incorporated in the mean-field models \citep[e.g.,][]{Munoz-Jaramillo_et_al_2009}.

The $\alpha$-effect is most important for the regeneration of poloidal
(north-south) magnetic field; mean toroidal fields can be generated
from a mean poloidal field by the shear of differential rotation in
what is generally called an ``$\Omega$-effect.''  These $\alpha\Omega$
dynamos and their variants have become central to the language used to
discuss solar and stellar dynamos.  Some live in the convection zone
alone while others rely on the interface at the base of the convection
zone, the tachocline, to generate cyclic reversals of global-scale
polarity.

Many recent solar dynamo models have also emphasized the meridional
circulation as a potentially important factor in promoting cyclic
magnetic activity
\citep[e.g.,][]{Wang&Sheeley_1991,Dikpati&Charbonneau_1999,Kuker_et_al_2001,
Nandy&Choudhuri_2001,Dikpati&Gilman_2006,Rempel_2006,
Jouve&Brun_2007,Yeates_et_al_2008}.  In these {\em Flux-Transport}
models, the equatorward migration of emerging bipolar active regions over the
course of the solar cycle is attributed to the equatorward advection
of toroidal flux in the lower convection zone by the mean circulation.
Many Flux-Transport models are also {\em Babcock-Leighton} models
whereby the principle source of mean poloidal field generation is the
buoyant emergence and subsequent dispersal of fibril toroidal flux
concentrations, often modeled as a non-local $\alpha$-effect
\citep{Charbonneau_2010}.  Although the physical origin of the
Babcock-Leighton mechanism is distinct from the turbulent
$\alpha$-effect underlying distributed and interface dynamo models,
the fields generated are still helical in nature and this helicity
ultimately arises from the rotation of the star.

The review by Charbonneau~\citep{Charbonneau_2010} is an excellent
place to read in more detail about these models, while
\cite{Brandenburg&Subramanian_2005} delves into the details of the
turbulence models themselves.  Comparisons between many of the codes
used in different mean-field models have been undertaken by
\citep{Jouve_et_al_2008}. Mean-field models still do not unambiguously
reproduce the solar cycle, despite heroic efforts leading up to the
current cycle \citep[e.g.,][]{Dikpati&Gilman_2006}.  To be clear,
these models are much further along than the 3D models, which are
beginning to regularly produce cyclic solutions only in recent time
\citep[e.g.,][]{Brown_et_al_2010b, Ghizaru_et_al_2010,
Kapyla_et_al_2010, Mitra_et_al_2010}.  Mean-field models can be run
for many solar cycles and are an invaluable tool for exploring the
extensive parameter space of solar and stellar convection, and several
explorations have been made of global-scale circulations in 
solar-like stars \citep[e.g.,][]{Rudiger_et_al_1998,Kuker&Stix_2001,
  Kuker&Rudiger_2005_A&A,  Kuker&Rudiger_2005_AN, Kuker&Rudiger_2008}.
However, many of the important underlying variables are difficult to constrain (e.g.,
the dependence of $\alpha$ on radius and latitude in the Sun; how
$\alpha$ varies with stellar properties such as rotation rate and
mass; the applicability of $\alpha$-effects to modeling the turbulent
induction observed in 3D models; etc.).  We turn now to a discussion
of fully non-linear 3D convection driven dynamos, which are beginning
to provide the opportunity to better constrain these unknown
quantities.

\begin{figure}[t]
\vspace{-0.5cm}
\includegraphics[width=\linewidth]{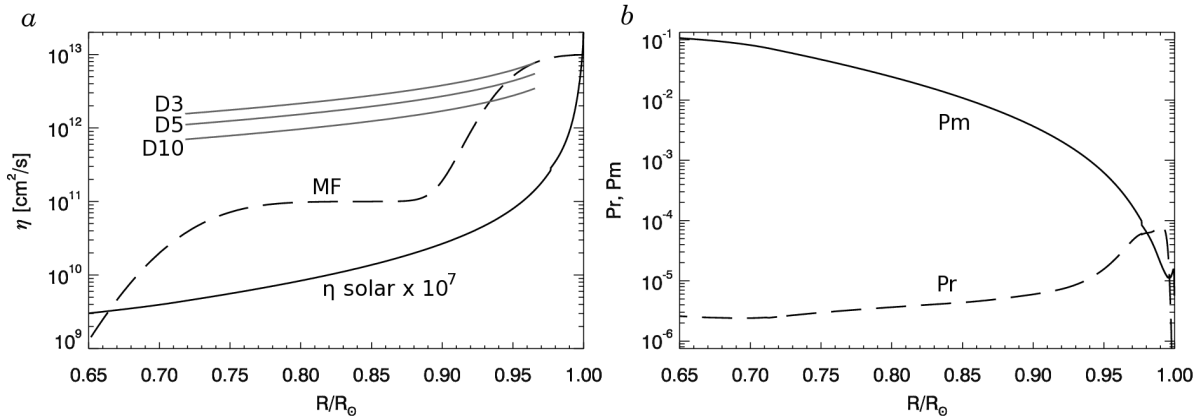}
\vspace{-1cm}
\caption{\label{fig:eta profiles} $(a)$ Radial profiles
    of magnetic diffusivity $\eta$ used in various models.  
    Shown is the double-step
    profiles used in the Babcock-Leighton models of
    \cite{Munoz-Jaramillo_et_al_2009} and similar to that used in
    \citep{Dikpati_et_al_2006, Dikpati&Gilman_2006} (MF,$\dashed$).  
    Three profiles from 3D MHD ASH dynamos are shown
    for comparison; these correspond to cases D3, D5 and D10 (grey
    lines).  The molecular diffusivity for a hydrogen plasma at
    solar conditions is also shown, multiplied by $10^7$ for display
    purposes.  $(b)$ The Prandtl number $\mathrm{Pr}=\nu/\kappa$ (dashed) and
    magnetic Prandtl number $\mathrm{Pm}=\nu/\eta$ (solid) for a
    hydrogen plasma at solar conditions.  Simulations use values of order unity.
  }
\end{figure}

\section{The gap between simulations and the Sun}

Numerical studies of the solar dynamo have a rich history, with the
the first 3D magnetohydrodynamic (MHD) convective global-scale solar
dynamo simulations attaining cyclic behavior in the early studies of
Gilman~\citep{Gilman_1983}.   Those Boussinesq simulations were quickly
joined by fully non-linear global-scale anelastic simulations, which
captured the stratified nature of the solar convection zone as well
\citep{Glatzmaier_1984, Glatzmaier_1985}.   Computational resources have grown at a
tremendous rate, and with them the complexities of the models studied.   
Modern simulations typically have higher resolutions and evolve for
longer intervals of time; in one simulation we will examine later
(case~D5), this represents roughly a factor of a million more 
computation than was possible in early studies \citep{Gilman_1983}.  This is is in
surprisingly good agreement with Moore's law doubling over the almost
thirty year interval separating these simulations, but it helps
clarify the huge gap remaining between solar convection and the
highest resolution simulations: another century of growth might
provide the resources to directly simulate solar convection on
global-scales.  

Stellar convection spans a vast range of spatial and temporal scales
which remain well beyond the grasp of direct numerical simulation.
Models of stellar dynamos must make various tradeoffs, either
building up from the diffusive scales or building down from the
global-scales. These are respectively called \emph{local} or
\emph{global} simulations; the latter will be our focus here.  
The highest resolution modern 3D simulations, running on
massively parallel supercomputers, capture roughly $1000^3$ total
points and typically can evolve for some $10^6-10^7$ timesteps.
In the Sun, the smallest scales of motion are set by diffusion and are
likely of order 1~mm \citep[\mbox{e.g., review}][]{Rieutord&Rincon_2010} while
the largest scales or motion are comparable to the solar radius
(700~Mm), with a total spectral range of almost $10^{12}$ in each of
three dimensions.  Temporal separations are similar, with fast
granulation on the surface overturning on roughly five minute
timescales while the deep structure of the Sun evolves over a span of gigayears.  

Clearly, stellar dynamo studies must drastically
simplify the physics of the stellar interior.  Molecular values of the
magnetic diffusivity $\eta$ for a hydrogen plasma under 
the conditions of the the solar convection zone range from roughly
$10^{2}$--$10^{6}\:\mathrm{cm}^2/\mathrm{s}$ as one moves from
the tachocline to the near surface regions, while the molecular
viscosity $\nu$ is of order $10^{\phantom{5}}\mathrm{cm}^2/\mathrm{s}$
in the solar convection zone \citep[e.g.,][]{Braginskii_1965, Rieutord_2008}.    These diffusivities are vastly smaller than the values used
in either mean-field models or 3D MHD simulations, and we illustrate
this in Figure~\ref{fig:eta profiles}. 

In contrast to the solar values, simulations typically employ values of
$\eta$ and $\nu$ that are of order $10^{12}\:\mathrm{cm}^2/\mathrm{s}$; this 
large value is more similar to simple estimates of turbulent diffusion
associated with granulation at the surface where 
$\nu_\mathrm{t} \sim V_\mathrm{t} L_\mathrm{t} \sim 10^{11}\:\mathrm{cm}^2/\mathrm{s}$ 
given $V_\mathrm{t} \sim 1\:\mathrm{km}/\mathrm{s}$ and $L_\mathrm{t}
\sim 1\:\mathrm{Mm}$ \citep[e.g.,][]{Rieutord&Rincon_2010}.  
Shown in Figure~\ref{fig:eta profiles}$a$ are radial profiles of
$\eta$ for both mean-field models  and 3D MHD dynamo simulations with
the ASH code.  Mean-field models often match to a turbulent diffusion
consistent with supergranulation at the surface and then taper to a
lower value at mid-convection zone (here $10^{11}\:\mathrm{cm}^2/\mathrm{s}$)
\citep[e.g.,][]{Dikpati_et_al_2006, Dikpati&Gilman_2006, Munoz-Jaramillo_et_al_2009}.  
Below the tachocline, $\eta_\mathrm{MF}$ is tapered further, sometimes
approaching the molecular values.  In 3D models the choice is often
made to scale $\eta$ with the background density.  Here a scaling of
$\rho^{-0.5}$ is used, though in other studies the exponent can scale
from 0 to -1 \citep[e.g.,][]{Brun_et_al_2004}.  The relative mixing
from diffusion of vorticity (by~viscosity~$\nu$), temperature
($\kappa$) and magnetism ($\eta$) is given by the Prandtl number and
magnetic Prandtl number, which are shown in Figure~\ref{fig:eta  profiles}$b$.
Molecular ratios for a hydrogen solar plasma are tiny, with Pm ranging from
$10^{-1}$ to $10^{-5}$ while Pr is of order $10^{-5}$.  Turbulent
values are likely of order unity but are not well constrained under
solar conditions; simulations typically take Pm and Pr to be near
unity as resolving large separations in diffusive scales requires very
high resolutions.

\section{Convection driven dynamos: the Sun}

Despite this daunting separation in parameter space, modern models are
making tremendous strides in understanding the non-linear couplings
between convection and rotation that build the solar differential
rotation.  It is now possible to study relatively high Reynolds number
convection (fluctuating Re of order a few hundred) in stratified
convection zones capturing density contrasts exceeding 100 (e.g., more
than 5 density scale heights) between the base of the convection zone
and the near-surface layers.
The anelastic spherical harmonic (ASH) code has been a very
useful tool in global-scale studies \citep{Clune_et_al_1999, 
Miesch_et_al_2000, Brun_et_al_2004}; tremendous progress has been made
in models of photospheric convection as well, but those will not be
the focus here \citep[e.g.,][]{Nordlund_et_al_2009}.

Simulations of solar convection self-consistently produce solar-like
profiles of differential rotation, with prograde equators, retrograde polar
regions and a monotonic decrease of angular velocity with latitude.
This profile is achieved partly through the redistribution of angular
momentum by turbulent Reynolds stresses in the convection
\citep{Brun&Toomre_2002, Miesch_2005, Miesch_et_al_2008}.  The
differential rotation profile is in what is called a ``thermal-wind
balance'', with an accompanying latitudinal gradient of temperature.
This is an effect that is well known in the geophysical community,
entering the vorticity evolution equation as a baroclinic term.
In solar convection, thermal-wind balance leads to more conical
profiles of angular velocity $\Omega$ \citep{Brun&Toomre_2002,
  Miesch_et_al_2008}.  Simulations suggest that in the
Sun the accompanying temperature perturbations at the solar surface
may be of order 1-10K.  Perturbations of this size at the base of the
convection zone, consistent with the geostrophic balances likely achieved in the
tachocline, can serve to tilt the contours of constant $\Omega$ until
very good agreement is attained between simulations and helioseismic
measurements \citep{Rempel_2005, Miesch_et_al_2006}.

Solar dynamo simulations generally produce complex magnetic  
topologies, with more than 95\% of the magnetic
energy in the fluctuating (non-axisymmetric) field
components \citep{Brun_et_al_2004}.  Mean fields are
complex, with multipolar structure and transient
toroidal ribbons and sheets.  Polarity reversals
of the dipole component occur but they are 
irregular in time.  The presence of an overshoot region and a tachocline 
of rotational shear promotes mean-field generation, 
producing persistent bands of toroidal flux antisymmetric 
about the equator while strengthening and stabilizing the dipole 
moment \citep{Browning_et_al_2006,Browning_et_al_2007,Miesch_et_al_2009}.
These simulations exhibit notable self-organization 
through the turbulent pumping of magnetic flux into 
the tachocline, amplification by rotational shear,
and selective shear-induced dissipation of small-scale
magnetic fluctuations.  Modern simulations build organized magnetic
fields and attain cyclic behavior, sometimes with a tachocline playing
an important role \citep{Ghizaru_et_al_2010, Kapyla_et_al_2010}, and
sometimes in the convection zone of the Sun alone
\citep{Miesch_et_al_2010, Mitra_et_al_2010}.  We now turn to the special
class of wreath-building convection zone dynamos.

\begin{figure}
\vspace{-0.3cm}
 \begin{center}
   \includegraphics{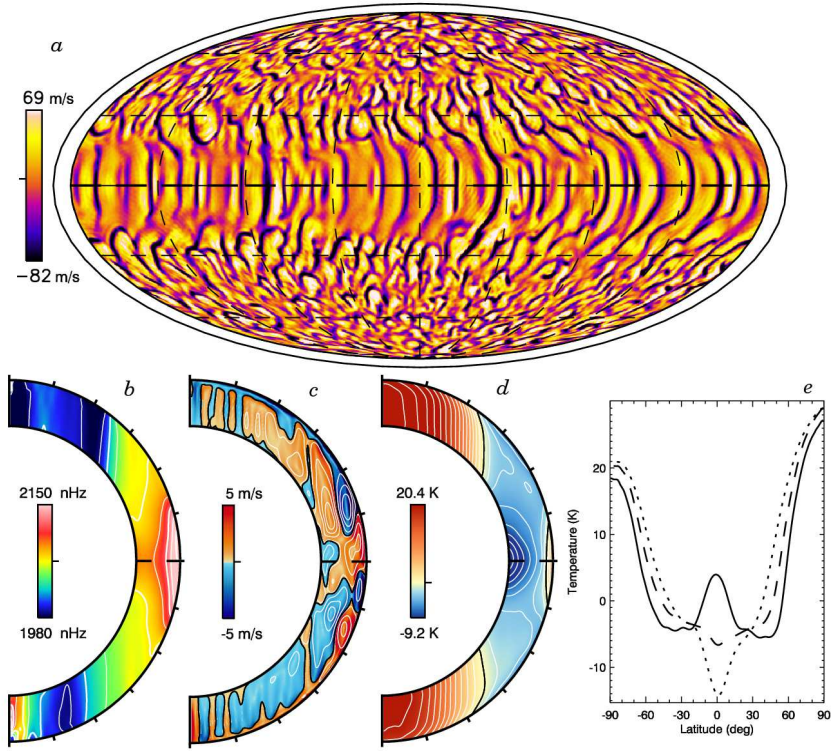}
 \end{center}
 \vspace{-0.75cm}
 \caption{Convection and global-scale flows in case~D5
   \citep{Brown_et_al_2010b}.  
   $(a)$ Convective patterns of radial velocity $v_r$ in global Mollweide
   projection at $0.95\:R_\odot$ with upflows light and downflows
   dark.  Poles are at top and bottom and thin line denotes stellar
   surface.  $(b)$~Differential rotation shown with longitudinally
   averaged angular velocity $\Omega$.  The rotation is solar-like,
   with fast (prograde) equator.  $(c)$~Meridional circulations
   with color indicating amplitude and sense of circulation (red
   counter-clockwise; blue clockwise) and mass flux streamlines
   overlaid.  Compared to solar simulations, the circulations here
   are broken into several weaker cells in both radius and latitude.
   $(d)$~Profile of mean temperature fluctuations.  This
   profile, with hot poles and cool mid-latitudes, represents the
   thermal wind balance achieved with the differential rotation.
   $(e)$~Latitude cuts of temperature at fixed radius, sampling top
   ($\full$), middle ($\broken$) and bottom ($\dotted$) of
   convection zone.  At the surface, the temperature contrast in
   latitude can reach $30$K.  Profiles shown in $b$--$e$ have been
   averaged in time over a 225 day interval.
   \label{fig:flows D5}
 }
\end{figure}

\section{Convection driven dynamos: rapidly rotating suns}
When stars like the Sun are younger they rotate much more rapidly.
These stars are observed to have strong surface magnetic activity and
are thought to have very active dynamos in their convection zones.
Rotational constraints are stronger in more rapidly rotating systems
and this can lead to greater correlations as convective structures
align with the rotation axis.  

Patterns of convection in a simulation of a young, rapidly rotating sun are shown in
Figure~\ref{fig:flows D5}$a$ near the stellar surface.  Convection
fills the domain and near the equator is strongly north-south aligned.
Correlations in these ``bannana-cells'' transport angular momentum and
build the profile of differential rotation shown in Figure~\ref{fig:flows D5}$b$; the
equator is fast, the poles are slow and the angular velocity contrast
is larger than in the Sun.  The meridional circulations in contrast
are weak and multi-celled in both radius and latitude
(Fig.~\ref{fig:flows D5}$c$).  Accompanying the angular velocity
profile is a large latitudinal gradient of temperature, shown
Figures~\ref{fig:flows D5}$d,e$.  Near the surface, there can be 30K 
contrasts between the hotter poles and cooler mid-latitudes.

\begin{figure}[!t]
\vspace{-0.2cm}
\begin{center}
  \includegraphics[width=\linewidth]{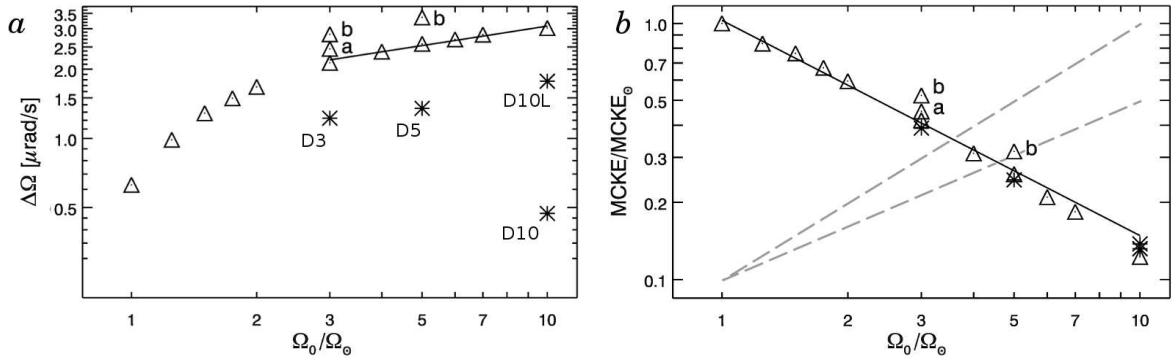}
\end{center}
\vspace{-0.75cm}
\caption{Global-scale flows and stellar rotation
  \citep{Brown_et_al_2008}.  ($a$) Angular 
  velocity shear of differential rotation $\Delta \Omega$ in
  latitude near the stellar surface shown as a function of rotation
  rate $\Omega_0$ relative to the solar rotation rate
  $\Omega_\odot$.  Hydrodynamic cases are shown with diamonds while
  dynamos are labeled and shown with asterisks.  $\Delta \Omega$
  grows with more rapid rotation in hydrodynamic cases.  Cases labeled
  a and b sample more turbulent states.  ($b$)
  Kinetic energy of the meridional circulations MCKE, normalized to
  that energy in the simulation at the solar rate.  MCKE decreases
  with more rapid rotation; a powerlaw of $\Omega_0^{-0.9}$ is shown
  for reference.  Grey dashed lines indicate scalings typically used
  in mean-field models for more rapidly rotating suns (e.g.,
  \cite{Charbonneau&Saar_2001, Dikpati_et_al_2001, Nandy_2004}, and
  see \cite{Jouve_et_al_2010} for models that follow the
  $\Omega_0^{-0.9}$ scaling).
  \label{fig:global-scale flows}
}
\end{figure}

The angular velocity contrast of the differential rotation $\Delta
\Omega$ and the kinetic energy contained in the meridional circulations (MCKE)
is shown for many rapidly rotating suns in Figure~\ref{fig:global-scale flows}.
Generally, we find that $\Delta \Omega$ grows with more rapid
rotation, while MCKE drops strongly \citep{Brown_et_al_2008}.  The
decrease of MCKE is a surprise and may hold important implications for
flux-transport dynamos in the mean-field framework \citep{Jouve_et_al_2010}. 
The growth of $\Delta \Omega$ is roughly in agreement with observations of
surface differential rotation in other stars, though substantial
disagreement remains between different groups of observers
\citep[e.g.,][]{Donahue_et_al_1996,Barnes_et_al_2005}. 
In hydrodynamic cases (triangles) this shear continues to
grow with faster rotation; in dynamo cases it may begin to saturate as
Lorentz forces become important and react back on the differential
rotation (asterisks).  In all cases, the growth of $\Delta \Omega$ with
rotation rate $\Omega_0$ is accompanied by a growing latitudinal
temperature contrast.  The temperature contrast grows from a few K at
the solar rate to several hundred K at the fastest rotation rates.

\begin{figure}[!p]
  \vspace{-0.3cm}
  \includegraphics[width=\linewidth]{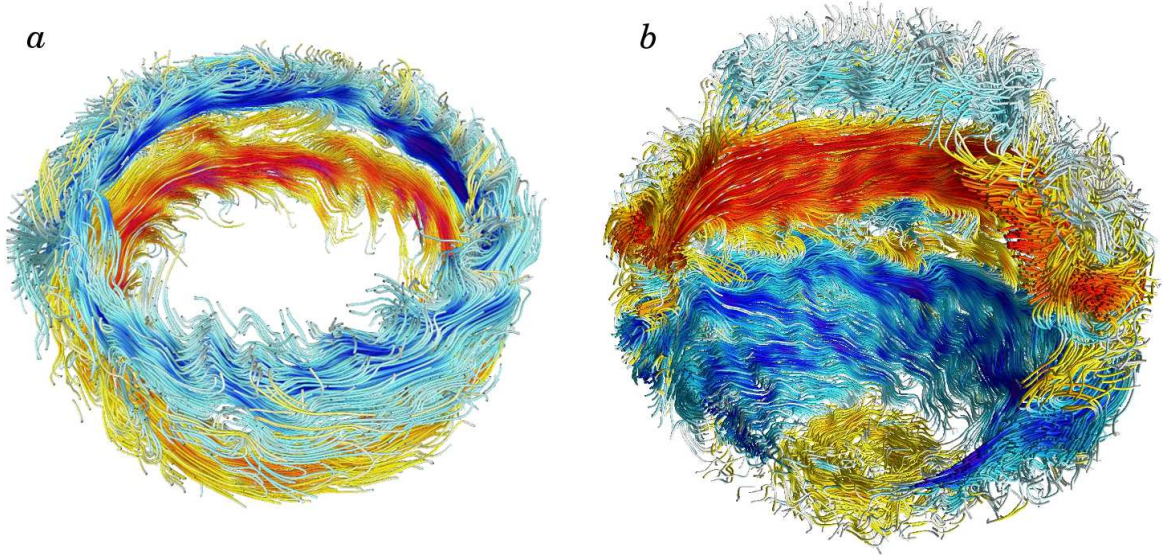}
  \caption{Magnetic wreaths in stellar convection zones.  $(a)$
    Persistent wreaths in case~D3.  Two wreaths of opposite polarity
    (red, positive; blue, negative) form above and below the equator.
    These magnetic structures coexist with the turbulent convection
    and retain their identity for more than 20,000 simulated days.
    $(b)$ Magnetic wreaths in cyclic case D5.  In this simulation the
    wreaths undergo reversals of polarity on roughly a 1500-day
    timescale.  Relic wreaths from the previous cycle are visible in
    the polar caps.  This snapshot is at same instant as
    Figure~\ref{fig:flows D5}$a$.
    \label{fig:magnetic wreaths D3 D5}
  }

\end{figure}

The magnetic fields created in dynamo simulations of rapidly rotating
suns are organized on global-scales into banded wreath-like
structures. These are shown for a dynamo at three times the solar
rotation rate (case~D3) in Figure~\ref{fig:magnetic wreaths D3 D5}$a$
\citep{Brown_et_al_2007c, Brown_et_al_2010a}. Two such wreaths are visible in the equatorial region, spanning the
depth of the convection zone and latitudes from roughly $\pm30^\circ$.
The dominant component of the magnetism is the longitudinal field
$B_\phi$, and the two wreaths have opposite polarities (red, positive;
blue, negative).  An even more rapidly rotating dynamo (case~D5) is
shown in Figure~\ref{fig:magnetic wreaths D3 D5}$b$.  Now the wreaths
fill the convection zone and the polar caps.  These wreaths show
significant time-variation and undergo quasi-regular polarity
reversals \citep{Brown_et_al_2010b}.

\begin{figure}[!p]
  \includegraphics[width=\linewidth]{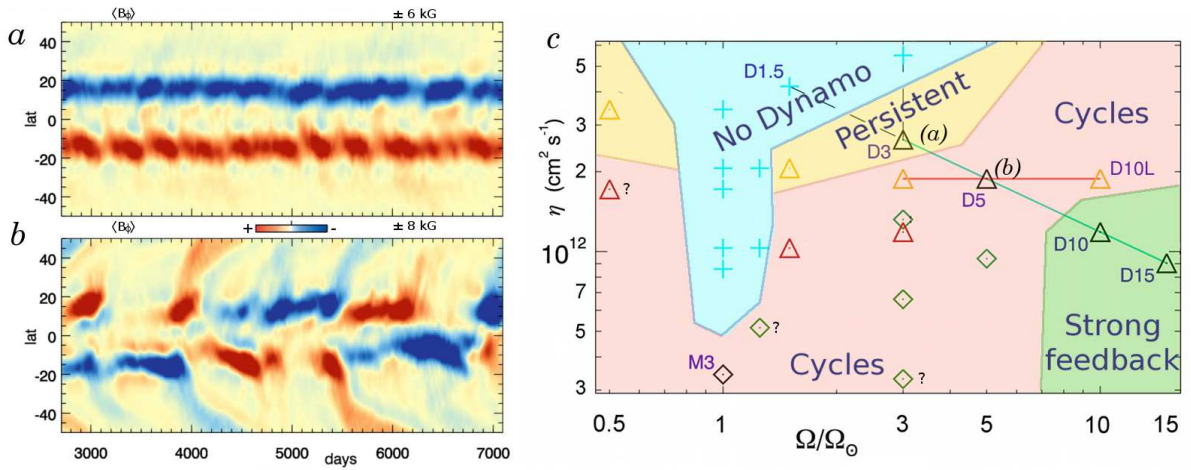}
  \caption{Family of wreath-building dynamo solutions.  
   $(a)$~Time-latitude plot of mean (axisymmetric) longitudinal
   magnetic field $B_\phi$ at mid-convection zone in persistent case D3 \citep{Brown_et_al_2010a}.   
   $(b)$~Cyclic case D5 shown for same span of time
   \citep{Brown_et_al_2010b}.  Three reversals are visible here,
   occurring on roughly 1500 day periods.
    $(c)$ Primary control parameters for simulations, with magnetic 
   diffusivity $\eta$ and rotation rate $\Omega_0$ sampling
   $0.5$--$15\Omega_\odot$.  Very approximate dynamo regimes are shown.
   Time dependence emerges at higher magnetic Reynolds number (lower
   $\eta$) and many dynamos undergo repeated reversals of global-scale
   polarity.  Cases with question marks show significant
   time-variation but have not been computed for long enough to definitively
   establish cyclic behavior.  
   At the highest rotation rates the Lorentz force can
   substantially reduce the differential rotation, but dynamo action
   is still achieved.  
   \label{fig:parameter space}
   }
\end{figure}

The time history of these two cases is shown in
Figures~\ref{fig:parameter space}$a,b$ (case D3 and D5 respectively).
Here the mean (axisymmetric) longitudinal magnetic field $\langle
B_\phi \rangle$ is shown at mid-convection zone over an interval of
about 4000 days; both cases have a full history of roughly 20,000
days.  In~case~D3 we generally find little time variation in the axisymmetric
magnetic fields associated with the wreaths; small variations are
visible on a roughly 500 day timescale, but the two wreaths retain
their polarities for the entire interval, which is significantly longer
than the convective overturn time (roughly 10--30 days), the rotation
period (9.3 days), or the ohmic diffusion time (about 1300 days at
mid-convection zone).  We refer to the dynamos in this regime as
persistent wreath-builders.  Case D5 (Fig.~\ref{fig:parameter
  space}$b$) is very different.  Here global-scale polarity reversals
occur roughly every 1500 days.  Three such reversals are shown here.
The ohmic diffusion time in this simulation is about 1800 days, while
the rotation period is 5.6 days.  

These two simulations are part of a much larger family of
wreath-building dynamos, which are summarized in Figure~\ref{fig:parameter
  space}$c$.  Shown here are 26 simulations at rotation rates ranging
from $0.5\:\Omega_\odot$ to $15\:\Omega_\odot$.  Wreath-building
dynamos are achieved in most simulations (17), though a smaller number
do not successfully regenerate their mean poloidal  fields (9,
indicated with crosses).    At individual rotation
rates (e.g., $3\:\Omega_\odot$), further simulations explore the
effects of lower magnetic diffusivity $\eta$ and hence higher magnetic
Reynolds numbers.  Some of these follow a path where the magnetic
Prandtl number Pm is fixed at 0.5 (triangles) while others sample up
to Pm=4 (diamonds).  Near the onset of dynamo action the wreaths
are similar to those found in case~D3 and persist for long intervals
with little variation in time.  At higher magnetic Reynolds numbers
(lower $\eta$ and higher $\Omega_0$) we find many simulations that
show cyclic reversals of global-scale magnetism (as in case~D5).  It
is difficult to determine what sets the cycle period in these dynamos:    
cycles appear to become shorter as $\eta$ decreases,
opposite to what might be expected if the ohmic time determined the
cycle period.  The dependence of cycle period on $\Omega_0$ is less
certain.  It is striking that coherent magnetic structures can arise
at all in the midst of turbulent convection.  We find the combination
of global-scale spatial organization and cyclic behavior fascinating,
as these appear to be the first self-consistent 3D convective stellar
dynamos to achieve such behavior in the bulk of the convection zone
rather than relying on a stable tachocline of shear.

\section{Where we now stand}

This is an exciting time in solar and stellar dynamo theory.
Stellar dynamo models have progressed tremendously in the past
decade.  Mean-field models are reaching a point where credible
predictions of upcoming solar cycles can be attempted.  Meanwhile,
several different 3D simulations using distinctly different
codes and assumptions have achieved global-scale 
magnetism and cyclic behavior in simulations of the solar dynamo
\citep[e.g.,][]{Ghizaru_et_al_2010,  Kapyla_et_al_2010,
  Miesch_et_al_2010, Mitra_et_al_2010, Nelson_et_al_2010}.  
Global-scale organization and cyclic reversals
are being found even in simulations without tachoclines. 
A major challenge now is to understand why
such cycles occur.  Significant progress can be made on this problem
by translating the results of 3D dynamo models into the language of
mean-field theory, measuring difficult to constrain quantities such as
$\alpha$ and the turbulent electromotive force (emf) that builds the
mean poloidal fields.  First attempts at diagnosing these quantities
are underway \citep[e.g.,][]{Brown_et_al_2010a, Brown_et_al_2010b,
  Racine_et_al_2010} but now 3D and mean-field modellers must partner
to better understand the cyclic dynamo 
simulations \citep[e.g.,][]{Miesch_2008, Jouve_et_al_2010}.  Such efforts will
refine the mean-field models but will also yield crucial insights into
the operation of the dynamos within the highly turbulent and
time-dependent 3D simulations. 

It is also crucial that further observational constraints be provided
for such 3D models; here is a brief and biased wish list.  In the
Sun, some sense of the deep meridional 
circulations would greatly enhance our confidence in modeling
results.  In particular, it would be useful to determine whether these
circulations are multi-cellular in latitude and radius, or whether one
large cell extends from the surface to the tachocline.  Simulations
generally find that thermal-wind balances arise along with the
differential rotation.  The amplitude of thermal perturbations at the
surface is likely to be quite small, of order 1-10K, but the detection
(or non-detection) of such a latitudinal gradient would be very
useful.  Estimates are beginning to be made based on simulations
\citep[e.g.,][]{Miesch_et_al_2006, Miesch_et_al_2008,
  Balbus_et_al_2009, Balbus&Latter_2010, Brun_et_al_2010} and attempts
have been made to observe this profile in the Sun \citep[e.g.,][and
references therein]{Rast_et_al_2008}.  Any estimate of the properties
of giant cell convection in the Sun would be tremendously useful as
well.  Direct detection of these structures would of course be ideal,
but much could be learned from indirect observations as well
\citep[e.g.,][]{Hanasoge_et_al_2010}. 

Convective dynamo models in 3D are being applied to other stars, 
both similar to and different from the Sun.  For the solar-type
stars, observations of the surface differential rotation $\Delta
\Omega$,  its variation with stellar mass and rotation rate, and its
temporal variation would greatly constrain these models.
Additionally, it is crucial to know how basic dynamo properties, such
as the amount of surface magnetism and the cycle period, scale with
differential rotation $\Delta \Omega$ rather than overall rotation
rate $\Omega_0$.  The thermal wind balances achieved in more rapidly
rotating solar-type stars likely lead to latitudinal temperature
contrasts of several hundred~K; in more massive and luminous F-type
stars, these contrasts may be as large as several thousand~K
(K.~Augustson, private communication).  These signatures may be observable.

\vspace{0.5cm} 
The research on wreath-building dynamos in rapidly rotating suns has
been done in collaboration with Matthew K.\ Browning, Allan Sacha
Brun, Mark S.\ Miesch, Nicholas J.\ Nelson and Juri Toomre.  I owe
them a great debt of gratitude.  I also thank Ellen Zweibel and Cary
Forest for inspiring the discussion of $\eta$ and plasma transport.
Funding for this research is provided in part through NSF Astronomy
and Astrophysics postdoctoral fellowship \mbox{AST 09-02004}. CMSO is
supported by NSF grant PHY 08-21899.  The simulations were carried out
with NSF PACI support of NICS, PSC, SDSC, and TACC.

\newcommand\aap{{A\&A}}%
\newcommand\apj{{ApJ}}%
\newcommand\apjl{{ApJ}}%
\newcommand\apjs{{ApJS}}%
\newcommand\jfm{{J. Fluid Mech.}}
\newcommand\an{{Astron. Nachr.}}
\newcommand\solphys{{Sol.~Phys.}}%
\newcommand\aapr{{A\&A~Rev.}}%
\newcommand\physrep{{Phys.~Rep.}}%
\newcommand\mnras{{MNRAS}}%

\bibliographystyle{iopart-num}
\begin{small}
\bibliography{bibliography}
\end{small}

\end{document}